\documentstyle[twocolumn]{article}
\newcommand{\Rset}[0]{\mbox{\rm I\kern-.200em R}}
\title{
NONLINEAR WAVE EQUATIONS FOR RELATIVITY}
\author{Maurice H.P.M. van Putten\\
	CRSR \& Cornell Theory Center\\
        Cornell University, Ithaca, NY 14853-6801, and\\
	Douglas M. Eardley,\\
	I.T.P., UCSB, Santa Barbara, CA 93106}
\date{}

\begin{document}

\maketitle

\begin{abstract}
Graviational radiation is described by canonical Yang-Mills wave equations
on the curved space-time manifold, together with evolution equations
for the metric in the tangent bundle.
The initial data problem is described in
Yang-Mills scalar and vector potentials, resulting in
Lie-constraints in addition to the familiar
Gauss-Codacci relations.
\end{abstract}

\bibliographystyle{plain}

\pagestyle{myheadings}
\markright{van Putten $et.$ $al.$}

\section{Introduction and outline of the approach}

The asymptotic gravitational wave structure resulting from
the coalescence of astrophysical black holes or
neutron stars is receiving wide attention in
connection with
the gravitational wave detectors currently
under construction. A compact binary system
produces strongly nonlinear gravitational waves
of which a small residue escapes to
infinity as radiation. The asymptotic wave structure
at a distant observer is in quantitative
relation to the
system parameters of compact binaries, which makes
gravitational radiation a new spectrum for
astronomical observations.
The prediction of gravitational wave forms is presently pursued
by numerical simulation. The current approaches are
based the 3+1 Hamiltonian formulation by Arnowitt, Deser and
Misner \cite{adm:a}, with the notable exception of strategies employing
null-coordinates ($e.g.$ \cite{cd:a})
or systems of conservation laws \cite{bm:a}.

The significance of gravity waves
in compact binaries, both in
the process of coalescence and
in radiation, suggests to focus
on a description of relativity by nonlinear wave equations.
Wave equations provide a proper frame-work for the physics
of nonlinear wave motion,
establish connection
with electromagnetics (abelian and linear) and Yang-Mills theory
(non-abelian and nonlinear; see, $e.g.$,
\cite{traut:a})
and may complement present approaches in numerical relativity
by way of understanding or circumventing
some of the computational difficulties.
This paper
establishes a
first step in this direction.
Generally, wave equations provide a setting appropriate for
numerical implementation via cylindrical or spherical
coordinate systems, including outgoing boundary conditions
at grid boundaries, and offer
a possible setting for numerical treatment
of ingoing horizon boundary conditions.

Pirani \cite{pirani:a} presented compelling arguments
to concentrate on the Riemann tensor
in describing gravitational radiation.
One can proceed by proposing to establish a numerical
algorithm based on the divergence of the Riemann tensor \cite{mvp:GR},
$R_{abcd}$, in which
the tensor
\begin{eqnarray}
\tau_{bcd}={16\pi}(\nabla_{[c}T_{d]b}-\frac{1}{2}g_{b[d}\nabla_{c]}T)
\label{enc}
\end{eqnarray}
acts as a source term.
A tetrad $\{(e_\mu)^b\}$-formalism takes the Riemannian
formulation into
equations of Yang-Mills type.
With Lorentz gauge
on the connection 1-forms, $\omega_{a\mu\nu}$,
\begin{eqnarray}
c_{\mu\nu}:=\nabla_c\omega^c_{\cdot\mu\nu}=0
\end{eqnarray}
($\nabla_a$ denotes the covariant derivative associated
with the metric, $g_{ab}$),
fully nonlinear, canonical Yang-Mills wave equations
are obtained
\begin{eqnarray}
\hat{{\Box}}\omega_{a\mu\nu}-R^c_a\omega_{c\mu\nu}
-[\omega^c,\nabla_a\omega_c]_{\mu\nu}
=\tau_{a\mu\nu}.
\end{eqnarray}
Here, $\hat{{\Box}}$ is the non-abelian generalization of the
Laplace-Beltrami wave operator.
The wave equations are also derived for
the scalar, Ricci rotation coefficients.
Complemented by the equations of structure for the
evolution of the tetrad legs, a complete system of evolution
equations is obtained which is amenable to numerical implementation.

This interwoveness of
wave motion and causal structure distinguishes gravity from
the other field theories.
In the present description, this two-fold nature of gravity has thus
been made explicit. Essentially, gravitational waves are now
propagated by wave equations on the (curved) manifold, while
the metric is evolved in the (flat) tangent bundle by the equations
of structure.
Of course, such two-fold description is only meaningful
for wave motion with wave lengths above the Planck scale, below which
the causal structure is not well-defined and wave motion can
not be distinguished from quantum fluctuations.

Because the wave equations are second order in time,
conditions on the initial data involve both $\omega_{a\mu\nu}$ and their
normal Lie derivates ${\cal L}_n\omega_{a\mu\nu}$.
In addition to the Gauss-Codacci relations,
$\omega_{a\mu\nu}$
and ${\cal L}_n\omega_{a\mu\nu}$
are related to the initial
energy-momentum and
$\tau_{b\mu\nu}$.
The latter follow from
{\em Gauss-Riemann relations},
as non-abelian generalizations of Gauss's law.

\section{Equations for $R_{abcd}$}

We will work on a four-dimensional manifold, $M$, with
hyperbolic metric $g_{ab}$. In a given
coordinate system $\{x^b\}$ the line-element is given by
\begin{eqnarray}
ds^2=g_{ab}\mbox{d}x^a\mbox{d}x^b.
\end{eqnarray}
The natural volume
element
on $M$ is $\epsilon_{abcd}=\sqrt{-g}\Delta_{abcd}$, where
$g$ denotes the determinant of the metric in the
given coordinate system, and $\Delta_{abcd}$
denotes the completely antisymmetric symbol.
Covariant differentiation associated with
$g_{ab}$ will be referred to by $\nabla_a$.
The geometry of $M$ is contained in the Riemann
tensor, $R_{abcd}$, which satisfies the
Bianchi identity
\begin{eqnarray}
\begin{array}{lll}
3\nabla_{[e}R_{ab]cd}&=&\nabla_eR_{abcd}+\nabla_aR_{bacd}\\
&&+\nabla_bR_{eacd}
=0.
\label{BI}
\end{array}
\end{eqnarray}
Using the volume element $\epsilon_{abcd}$,
the dual $*R$ is defined as
%$\frac{1}{4}\epsilon_{ab}^{\cdots ef}\epsilon_{cd}^{\cdots gh}
%R_{efgh}$.
$\frac{1}{2}\epsilon_{ab}^{\cdots ef}
R_{efcd}$. The Bianchi identity then takes the form
\begin{eqnarray}
\nabla^a*R_{abcd}=0.
\end{eqnarray}

Einstein
proposed that the
Ricci tensor $R_{ab}=R_{acb}^{\hskip0.15in c}$
and scalar curvature $R=R_c^{\cdot c}$ are connected to
energy-momentum,
$T_{ab}$, through
\begin{eqnarray}
R_{ab}-\frac{1}{2}g_{ab}R=8\pi T_{ab}.
\label{EINA}
\end{eqnarray}
Geometry now becomes dynamical with
(\ref{BI}) (see, $e.g.$, \cite{wg:g})
\begin{eqnarray}
\nabla^dR_{abcd}=2\nabla_{[b}R_{a]c}.
\label{div_R}
\end{eqnarray}
In the presence of Einsteins equations
(\ref{EINA}), therefore, (\ref{div_R}) becomes
\begin{eqnarray}
E:\nabla^{a}R_{abcd}=16\pi(\nabla_{[c}T_{d]b}-\frac{1}{2}g_{b[d}
\nabla_{c]}T).
\end{eqnarray}
The quantity on the right-hand side shall be referred to as
$\tau_{bcd}$.
In vacuo,  $E$
has been discussed by Klainerman \cite{klain:a},
who refers to $E$ (with $\tau_{bcd}=0$) together with (\ref{BI})
as the spin-2
equations.
The tensor $\tau_{bcd}$ is divergence free:
\begin{eqnarray}
\nabla^b\tau_{bcd}=0,
\end{eqnarray}
in consequence of the conservation laws
$\nabla^aT_{ab}=0$,
and in agreement with
$\nabla^b\nabla^aR_{abcd}=0.$

\subsection{Yang-Mills equations}

In the language of tetrads, additional invariance
arises due to the liberty of choosing the
tetrad position at each space-time point.
This invariance is described by the Lorentz group, and
introduces a vector gauge invariance in the form of the
connection 1-forms, governed by
Yang-Mills equations as outlined below.
The formal arguments can be found
in the theory of supersymmetry (see, $e.g.$, \cite{traut:a,muller:m,gieres:f}).

Let $\{(e_\mu)^b\}$ denote a tetrad satisfying
\begin{eqnarray}
\left\{
\begin{array}{ll}
(e_\mu)^c(e_\nu)_c&=\eta_{\mu\nu},\\
\eta^{\mu\nu}(e_\mu)_a(e_\nu)^b&=\delta_a^b,\\
\eta_{\mu\nu}&=\delta_{\mu\nu}\mbox{sign}(\eta_{\mu\nu}).
\end{array}
\right.
\label{tetrad}
\end{eqnarray}
Neighboring tetrads introduce their connection 1-forms, $\omega_{a\mu\nu}$,
\begin{eqnarray}
\omega_{a\mu\nu}:=(e_\mu)^c\nabla_a(e_\nu)_c.
\label{defom}
\end{eqnarray}
The connection 1-forms $\omega_{a\mu\nu}$ serve as
Yang-Mills connections in the gauge covariant derivative
\begin{eqnarray}
\hat{\nabla}_a=\nabla_a+[\omega_a,\cdot],
\end{eqnarray}
satisfying $\hat{\nabla}_a(e_\mu)^b=0$.
Here, the commutator is defined by its action on
tensors
$\phi_{a_{1}\cdots a_k\alpha_{1}\cdots\alpha_{l}}$
as
\begin{eqnarray}
\begin{array}{ll}
 [\omega_a,&\phi_{a_1\cdots a_k}]_{\alpha_1\cdots\alpha_l}\\&=
 \sum_i
 \omega_{a\alpha_i}^{\hskip.10in\alpha_j}
 \omega_{a_1\cdots a_k\alpha_1\cdots\alpha_j\cdots\alpha_l}.
\label{LB1}
\end{array}
\end{eqnarray}
In particular, we have
\begin{eqnarray}
 [\omega_a,\omega_b]_{\mu\nu}=
 \omega_{a\mu}^{\hskip.10in\alpha}\omega_{b\alpha\nu}
-\omega_{a\nu}^{\hskip.10in\alpha}\omega_{b\alpha\mu}.
\label{LB2}
\end{eqnarray}
In what follows, Greek indices stand for contractions with
tetrad elements: if $v^b$ is a vector field, then
$v_\mu=v^b(e_\mu)_b$, and $v^\mu=\eta^{\mu\nu}v_\nu$.

The Yang-Mills construction
thus obtains
for the Bianchi identity
\begin{eqnarray}
\hat{\nabla}^a*R_{ab\mu\nu}&=0,
\label{hatBI}
\end{eqnarray}
and for the equivalent of $E$
\begin{eqnarray}
E^\prime:
\hat{\nabla}^a R_{ab\mu\nu}&=\tau_{b\mu\nu}.
\label{YM}
\end{eqnarray}
The Bianchi identity (\ref{hatBI})
introduces the representation
($cf.$ \cite{wg:g})
\begin{eqnarray}
R_{ab\mu\nu}=\nabla_a\omega_{b\mu\nu}-\nabla_b\omega_{a\mu\nu}
+[\omega_a,\omega_b]_{\mu\nu},
\label{rep}
\end{eqnarray}
which will be central
in our discussion.

The antisymmetries in the Riemann tensor
introduce conditions on initial data
on an initial hypersurface, $\Sigma$. If $\nu^b$ denotes the normal to
$\Sigma$, and
\begin{eqnarray}
\nabla_a=-\nu_a(\nu^c\nabla_c)+D_a\mbox{  on  }\Sigma,
\label{denab}
\end{eqnarray}
we obtain the {\em Gauss-Riemann} relations
\begin{eqnarray}
\left\{
\begin{array}{ll}
\nu^bD^aR_{abcd}  &=-\rho_{cd},\\
\nu^bD^a*R_{abcd}&=0,
\end{array}
\label{GRI}
\right.
\end{eqnarray}
where $\rho_{cd}=\nu^b\tau_{bcd}$.
These may be considered generalizations
of Gauss's law in
electromagnetism.
Conditions
(\ref{GRI}) find their equivalents
in the tetrad formulation. To this end, we write,
analogous to (\ref{denab}), on $\Sigma$ the derivative as
\begin{eqnarray}
\hat{\nabla}_a=
-\nu_a(\nu^c\hat{\nabla}_c)
+\hat{D}_a.
\label{nablaS}
\end{eqnarray}
The Gauss-Riemann relations (\ref{GRI}) thus become
\begin{eqnarray}
\left\{
\begin{array}{ll}
\nu^b\hat{D}^aR_{ab\mu\nu}&=\rho_{\mu\nu},\\
\nu^b\hat{D}^a*R_{ab\mu\nu}&=0.
\end{array}
\right.
\label{GRII}
\end{eqnarray}

\subsection{Evolution equations for the tetrads}

The definition of the connection 1-forms gives
the equations of structure \cite{wg:g}
\begin{eqnarray}
\partial_{[a}(e_\mu)_{b]}=(e^\nu)_{[b}\omega_{a]\nu\mu}.
\label{EL}
\end{eqnarray}
In (\ref{EL}),
$\partial_t(e_\mu)_t$
is left undefined. Defining $\xi^b=(\partial_t)^b$,
the four time-components
\begin{eqnarray}
N_\mu:=(e_\mu)_a\xi^a
\label{LS}
\end{eqnarray}
become freely specifyable functions.
The evolution equations for the tetrad legs thus become
\begin{eqnarray}
\partial_t(e_\mu)_b+\omega_{t\mu}^{\hskip.10in\nu}(e_\nu)_b=
\partial_bN_\mu+\omega_{b\mu}^{\hskip.10in\nu}N_\nu.
\label{ELB}
\end{eqnarray}
The {\em tetrad lapse} functions $N_\mu$
are related to
the familiar lapse and shift functions
in the Hamiltonian formalism through
\begin{eqnarray}
g_{at}=N_\alpha(e^\alpha)_a
=(N_qN^q-N^2,N_p).
\end{eqnarray}

We will now turn to evolution equations for the
connection 1-forms.

\section{Equations for $\omega_{a\mu\nu}$}

We define a {\em Lorentzian} cross-section
of the tangent bundle of the space-time manifold by
\cite{mvp:GR}
\begin{eqnarray}
c_{\mu\nu}:=\nabla^d\omega_{d\mu\nu}=0.
\label{LT}
\end{eqnarray}
The {\em Lorentz gauge} (\ref{LT})
provides\footnote{Conceptually, $c_{\mu\nu}=f(\omega_{a\mu\nu},g_{ab})$
will also serve its purpose, where $f(\cdot,\cdot)$ depends analytically on its
arguments.}
a complete, six-fold connection between
neighboring tetrads.
The six {\em constraints} $c_{\mu\nu}=0$ are
incorporated in $E^\prime$ by application of the divergence
technique \cite{mvp:91,mvp:VIII}:
\begin{eqnarray}
E^{\prime\prime}:
\hat{\nabla}^a\{R_{ab\mu\nu}+g_{ab}c_{\mu\nu}\}=\tau_{b\mu\nu}.
\label{DIV}
\end{eqnarray}

Recall the transformation rule for the
connection (see, $e.g.$ \cite{gieres:f})
\begin{eqnarray}
\omega_{a\bar{\mu}\bar{\nu}}
=\Lambda_{\bar{\mu}}^{\hskip0.07in\alpha}
 \Lambda_{\bar{\nu}}^{\hskip0.07in\beta}
 \omega_{a\alpha\beta}
+\Lambda_{\bar{\mu}}^{\hskip0.07in\alpha}\nabla_a
 \Lambda_{\bar{\nu}\alpha}.
\label{rule}
\end{eqnarray}
In the present tetrad language, this gauge transformation is readily
established by consideration of two tetrads,
$\{(e_\mu)^b\}$
and $\{(\bar{e}_\mu)^b\}$. The construction
$\Lambda_{\bar{\mu}}^{\hskip0.1in \nu}:=
(\bar{e}_{\bar \mu})_c
(e^\nu)^c$
provides a finite transformation
$v_{\bar{\mu}}=\Lambda_{\bar{\mu}}^{\hskip0.1in\alpha}v_{\alpha}$
of a field $v_{\alpha}=(e_\alpha)^bv_b$
in the $\{(e_\mu)^b\}$-tetrad representation into $v_{\bar{\alpha}}$ in the
$\{(\bar{e}_\mu)^b\}$-tetrad representation. This applied to (\ref{defom})
obtains (\ref{rule}).

\subsection{Existence of Lorentzian cross-section}

To proceed, we consider in an open neighborhood
${\cal N}(\Sigma)$ of $\Sigma$ with
Gaussian normal coordinates
$\{\tau,x^p\}$ ($\nu^c\nabla_c\tau=1$) the infinitesimal
Lorentz transformation
\begin{eqnarray}
\Lambda_{\mu}^{\hskip0.08in\nu}=\delta_{\mu}^{\hskip0.08in\nu}
+\frac{1}{2}\tau^2\sigma_{\mu}^{\hskip0.08in\nu}
\mbox{  in  }{\cal N}(\Sigma).
\label{tLT}
\end{eqnarray}
It is convenient to employ language of
scalar and vector
potentials, $\Phi_{\mu\nu}$ and $A_{a\mu\nu}$, respectively, defined in
\begin{eqnarray}
\omega_{a\mu\nu}=\nu_a\Phi_{\mu\nu}+A_{a\mu\nu}.
\label{DEOM}
\end{eqnarray}
Denoting
the effect of (\ref{tLT}) via (\ref{rule}) by a superscript $(r)$,
we have
\begin{eqnarray}
\omega^{(r)}_{a\mu\nu}=\omega_{a\mu\nu}+\tau\delta_a^\tau\sigma_{\mu\nu}
\mbox{  in  }{\cal N}(\Sigma),
\label{rLT}
\end{eqnarray}
so that
\begin{eqnarray}
\Phi^{(r)}_{\mu\nu}=\Phi_{\mu\nu}
-\tau\sigma_{\mu\nu}
\mbox{  in  }{\cal N}(\Sigma).
\end{eqnarray}
Using a geodesic extension of the normal $\nu^b$ off $\Sigma$,
constraints (\ref{LT}) become
\begin{eqnarray}
c_{\mu\nu}
&=&\nabla_c\omega^c_{\cdot\mu\nu}
=\dot{\Phi}_{\mu\nu}+D_c\omega^c_{\cdot\mu\nu}\mbox{  on  }\Sigma.
\end{eqnarray}
In $A_{a\mu\nu}$, this obtains
\begin{eqnarray}
\begin{array}{ll}
c_{\mu\nu}
&=\dot{\Phi}_{\mu\nu}+D_c(\nu^c\Phi_{\mu\nu}+A^c_{\cdot\mu\nu})\\
&=\dot{\Phi}_{\mu\nu}+K\Phi_{\mu\nu}+D_cA^c_{\cdot\mu\nu}
\end{array}
\label{cB}
\end{eqnarray}
on $\Sigma.$
Consequently, we have the transformations
\begin{eqnarray}
\left.
\begin{array}{ll}
\Phi^{(r)}_{\mu\nu}&=\Phi_{\mu\nu}\\
\dot{\Phi}^{(r)}_{\mu\nu}
&=\dot{\Phi}_{\mu\nu}-\sigma_{\mu\nu}\\
A_{a\mu\nu}^{(r)}&=A_{a\mu\nu}
\end{array}
\right\}
\mbox{  on  }\Sigma.
\label{PhiLT}
\end{eqnarray}
By choice of $\sigma_{\mu\nu}=c_{\mu\nu}$,
the constraints (\ref{LT}) transform into
\begin{eqnarray}
c^{(r)}_{\mu\nu}=c_{\mu\nu}-\sigma_{\mu\nu}=0.
\label{LTZ}
\end{eqnarray}
Notice that bringing the tetrad in Lorentz gauge (\ref{LTZ})
by (\ref{tLT}) is achieved by proper second time-derivative
$\partial_\tau^2 (E_\mu)^b_{|\tau=0}$ of its legs,
leaving the tetrad position and its first
time-derivative as invariants at $\tau=0$.

We shall now establish that $E^{\prime\prime}$
maintains the Lorentz gauge (\ref{LT}) in the future domain
of dependence of $\Sigma$.
The inhomogeneous Gauss-Riemann relation (\ref{GRI})
is implied by antisymmetry of the Riemann tensor in its
coordinate indices, and gives
\begin{eqnarray}
\begin{array}{ll}
0&=\nu^b\{\hat{\nabla}^a(R_{ab\mu\nu}+g_{ab}c_{\mu\nu})-\tau_{b\mu\nu}\}\\
&=
\nu^b(\hat{D}^aR_{ab\mu\nu}-\tau_{b\mu\nu})+(\nu^b\hat{\nabla}_b)c_{\mu\nu}\\
&=(\nu^b\hat{\nabla}_b)c_{\mu\nu}.
\end{array}
\label{zc}
\end{eqnarray}
The inhomogeneous Gauss-Riemann relations are
gauge covariant, so that
we are at liberty to consider initial data
satisfying both (\ref{zc}) and the gauge choice (\ref{LTZ}), whence
\begin{eqnarray}
c_{\mu\nu}=
(\nu^c\hat{\nabla}_c)c_{\mu\nu}=0
\mbox{  on  }\Sigma.
\label{cIN}
\end{eqnarray}
In (\ref{DIV})
$c_{\mu\nu}$ satisfies a homogeneous
Yang-Mills wave equation
($cf.$ \cite{mvp:91,mvp:VIII}),
\begin{eqnarray}
\hat{{\Box}}c_{\mu\nu}:=\hat{\nabla}^c\hat{\nabla}_cc_{\mu\nu}=0.
\label{cWE}
\end{eqnarray}
It follows that the scalar fields $c_{\mu\nu}$ are solutions to
an initial value problem for homogeneous wave equations (\ref{cWE})
with trivial
Cauchy data (\ref{cIN}). Consequently,
\begin{eqnarray}
c_{\mu\nu}=0\mbox{  in  }D^+(\Sigma),
\label{c00}
\end{eqnarray}
which establishes that solutions to $E^{\prime\prime}$
%(\ref{DIV})
are solutions to
$E^\prime$.

We now elaborate further on (\ref{DIV}).
\subsection{Canonical wave equations}

In Lorentz gauge (\ref{LT}),
the divergence equation $E^{\prime\prime}$
obtains wave equations for the connection 1-forms
through the representation of the Riemann tensor (\ref{rep}).
Indeed, by explicit calculation, we have
\begin{eqnarray}
\hat{\Box}\omega_{a\mu\nu}-R_a^c\omega_{c\mu\nu}
-[\omega^c,\nabla_a\omega_c]_{\mu\nu}=\tau_{a\mu\nu}.
\label{OM}
\end{eqnarray}
Here, we have used
$c_{\mu\nu}=0$, so that
$\hat{\nabla}_ac_{\mu\nu}
=\nabla_ac_{\mu\nu}$.
${\hat{\Box}}$ is used to denote
the Yang-Mills wave operator $\hat{\nabla}^c\hat{\nabla}_c$.
The Ricci tensor $R_{ab}$ in (\ref{OM})
is understood in terms of $T_{ab}$
using Einstein's equations.

Similarly, we can obtain
a system of scalar equations for the Ricci rotation coefficients,
$\omega_{\alpha\mu\nu}=(e_\alpha)^a\omega_{a\mu\nu}$.
This involves a number of manipulations,
the result of which is
\begin{eqnarray}
\begin{array}{rl}
\hat{\Box}\omega_{\alpha\mu\nu}
-&R^\beta_\alpha\omega_{\beta\mu\nu}
-\omega_{\alpha}^{\hskip0.07in\gamma\beta}[\omega_\gamma,\omega_\beta]_{\mu\nu}\\
&-[\omega^\beta,\partial_\alpha\omega_\beta]_{\mu\nu}
=\tau_{\alpha\mu\nu}.
\label{scalar}
\end{array}
\end{eqnarray}

The two-fold nature of gravity can now be expressed as
\newtheorem{thm2}{Separation Theorem.}
\begin{thm2}
%[\mbox{Separation Theorem}]
Gravitational wave motion
is governed by canonical
wave equations on $M$.
In response to the wave motion,
the metric on $M$ evolves in the tangent bundle of
$M$ by the equations of structure.
\end{thm2}

The coupling of matter to the connections is accounted for by $\tau_{bcd}$.
It is of interest to note that $\tau_{bcd}$ contains vorticity;
a detailed enumeration of the nature of $\tau_{bcd}$
falls outside the scope of
this dicsussion.

The Theorem suggests some computational approximations
for weakly nonlinear gravity waves. Firstly,
consider equations with uncoupled (prescribed or fixed) metric,
with given
Laplace-Beltrami wave operator, $\Box_{g}$.
Thus,
we have (in vacuo)
\begin{eqnarray}
\hat{\Box}_{g}\omega_{a\mu\nu}-[\omega^c,\nabla_a\omega_c]_{\mu\nu}=0,
\end{eqnarray}
with the metric $g_{ab}$ as a background field providing the
underlying causal structure.
The approximate radiation then follows
from the fluctuations in the metric as would follow
from the
equations of structure.
Secondly, small amplitude waves allow us to linearize, thereby
obtaining the purely abelian wave equations
\begin{eqnarray}
\Box_{\eta}\omega_{a\mu\nu}=0,
\end{eqnarray}
where $\eta$ refers to the Lorentz metric.
Numerical experiments must show the validity of such
approximations.

\section{Initial value problem}

Initial data for the wave equations are
$\omega_{a\mu\nu}$
and its Lie derivative ${\cal L}_n\omega_{a\mu\nu}$ on
$\Sigma$.
These data must satisfy
certain constraints on $\Sigma$, commensurate with
the initial distribution of energy-momentum and the Gauss-Riemann
equations (\ref{GRII}). We shall express these equations in terms of
scalar and vector
potentials (\ref{DEOM}).

The projection tensor,
$h_{ab}$, onto $\Sigma$ is
\begin{eqnarray}
h_{ab}&=g_{ab}+\nu_a\nu_b.
\end{eqnarray}
The covariant derivative
induced by $h_{ab}$ shall be denoted by
$\bar{D}_a$, $i.e.$, $\bar{D}_ah_{cd}=0$.
In what follows,
$(\nu^c\nabla_c)f$
is also denoted by $\dot{f}$.
The unit normal is extended geodesically off $\Sigma$, so that
$\nu^c\nabla_c\nu^b=0$ and
$(\nu^c\nabla_c)h_{ab}=h_{ab}(\nu^c\nabla_c)$.
Thus, $\dot{\Phi}=(\nu^c\nabla_c)\Phi$
and $\dot{A}_{a\mu\nu}=(\nu^c\nabla_c)A_{a\mu\nu}$.
We further define
\begin{eqnarray}
\left\{
\begin{array}{ll}
\Phi^\prime&:=\dot{\Phi}_{\mu\nu}+K\Phi_{\mu\nu},\\
A^\prime_{a\mu\nu}
&:={\cal L}_nA_{a\mu\nu}
=\dot{A}_{a\mu\nu}+K_a^{\hskip.04in c}A_{c\mu\nu}.
\end{array}
\right.
\end{eqnarray}

\subsection{Expressions for $A_{a\mu\nu}$}

Consider a {\em normal tetrad},
$\{(E_\mu)^b\}$,
in which
one of the legs is (initially) everywhere normal to the initial hypersurface:
\begin{eqnarray}
(E_{\mu=n})^b=\nu^b\mbox{ on }\Sigma.
\label{NT}
\end{eqnarray}
Extrinsic curvature, $K_{ab}$, of
$\Sigma$ can be defined
`static' by
projection of the unit normal following
parallel transport over $\Sigma$,
\begin{eqnarray}
K_{ab}=\bar{D}_a\nu_b,
\label{KAB}
\end{eqnarray}
and `dynamic' by the Lie derivative, $\frac{1}{2}{\cal L}_nh_{ab}$,
of the projection
operator $h_{ab}$ with respect to a Gaussian normal
coordinate, $n$.
With a normal tetrad,
its (extrinsic) helicity, $i.e$, the twist in a strip
swept out by the integral curves of a leg $(E_\nu)^b$
passing through an integral curve of leg $(E_\mu)^b$
is
\begin{eqnarray}
H_{\mu\nu}:=\nu_b(E_\mu)^c\nabla_c(E_\nu)^b\mbox{  on  }\Sigma.
\end{eqnarray}
For $\mu,\nu\ne n$,
$H_{\mu\nu}$ describes twist in $\Sigma$
when $\mu\ne\nu$, and bending when $\mu=\nu$.
It follows that
\begin{eqnarray}
H_{\alpha\beta}h^\alpha_\mu h^\beta_\nu
=-K_{\mu\nu}
\label{HK}
\end{eqnarray}
in view of (\ref{NT}):
$\nu_a(E_\alpha)^a=-\delta_{\alpha}^{\nu}$ on $\Sigma$.
The normal ($\mu=n)$ and tangent
($\mu\ne n$)
extrinsic helicities
are now
\begin{eqnarray}
\begin{array}{ll}
H_{\mu\nu}&=\nu_b(E_\mu)^c\{\omega_{c\gamma\nu}(E^\gamma)^b\}\\
&=\left\{
\begin{array}{ll}
-\Phi_{n\nu}&\mbox{  if  }\mu=n,\\
A_{\mu n\nu}&\mbox{  if  }\mu\ne n.
\end{array}
\right.
\label{HA}
\end{array}
\end{eqnarray}
By (\ref{HK}),
the symmetry of $K_{ab}$ and $A_{\mu n \nu}=-A_{\mu\nu n}$,
we have for $\mu,\nu\ne n$ the symmetry
\begin{eqnarray}
A_{[\mu\nu]n}=0.
\end{eqnarray}
If $\mu\ne n$,
$(E_\mu)_a=h_{ab}(E_\mu)^b$, and so
\begin{eqnarray}
\begin{array}{ll}
A_{a\mu\nu}&=h^b_a(E_\mu)_c\nabla_b(E_\nu)^c\\
&=(E_\mu)_c[h^c_dh^b_a\nabla_b](E_\nu)^d\\
&\equiv(E_\mu)_c\bar{D}_a(E_\nu)^c.
\end{array}
\label{ex1}
\end{eqnarray}
If also $\nu\ne n$, $\bar{D}_a$ acts in (\ref{ex1}) on tangent legs
$(E_\nu)^b$,
the result of which is determined
by $h_{ab}$.
We shall write $\bar{A}_{a\mu\nu}=A_{a\alpha\beta}h^\alpha_\mu h^\beta_\nu$
for this intrinsic part of $A_{a\mu\nu}$.
In the normal tetrad, therefore, $A_{a\mu\nu}$ falls into two
groups: (i) the extrinsic part $A_{\mu\nu n}=K_{\mu\nu}$,
and the (ii) intrinsic part $\bar{A}_{a\mu\nu}$, with which we have
\begin{eqnarray}
%\begin{array}{ll}
A_{\alpha\mu\nu}
%&=\bar{A}_{\alpha\mu\nu}+A_{\alpha n\nu}\delta_{\mu}^n
%+A_{\alpha \mu n}\delta_{\nu}^n\\
&=\bar{A}_{\alpha\mu\nu}+2K_{\alpha[\mu}\delta^n_{\nu]}
\mbox{  on  }\Sigma.
\label{decA}
%\end{array}
\end{eqnarray}

\subsection{Expressions for $R_{ab\mu\nu}$}

On $\Sigma$, $R_{ab\mu\nu}$ contains
intrinsic, three-dimensional, and extrinsic
parts by substitution of (\ref{DEOM}):
\begin{eqnarray}
\begin{array}{rl}
R_{ab\mu\nu}
=&\nabla_a\omega_{b\mu\nu}-\nabla_b\omega_{a\mu\nu}
   +[\omega_a,\omega_b]_{\mu\nu}\\
%=&(-\nu_a(\nu^c\nabla_c)+D_a)
%    (\nu_b\Phi_{\mu\nu}+A_{b\mu\nu})\\
%  &-(-\nu_b(\nu^c\nabla_c)+D_b)(\nu_a\Phi_{\mu\nu}+A_{a\mu\nu})\\
%=&-\nu_a\nu_b\Phi_{\mu\nu}-\nu_a\dot{A}_{b\mu\nu}+K_{ab}\Phi_{\mu\nu}\\
%  &+\nu_bD_a\Phi_{\mu\nu}+D_aA_{b\mu\nu}
%    +\nu_b\nu_a\Phi_{\mu\nu}\\
%  &+\nu_b\dot{A}_{a\mu\nu}-K_{ba}\Phi_{\mu\nu}
%    -\nu_aD_b\Phi_{\mu\nu}\\
%  &-D_bA_{a\mu\nu}
%    +[\omega_a,\omega_b]_{\mu\nu}\\
%=&-\nu_a\dot{A}_{b\mu\nu}
%    +\nu_b\dot{A}_{a\mu\nu}
%    +\nu_bD_a\Phi_{\mu\nu}\\
%  &+D_aA_{b\mu\nu}
%    -\nu_aD_b\Phi_{\mu\nu}-D_bA_{a\mu\nu}\\
%  &+[\omega_a,\omega_b]_{\mu\nu}\\
=&2\nu_{[b}\hat{D}_{a]}\Phi_{\mu\nu}+2\nu_{[b}\dot{A}_{a]\mu\nu}\\
  &+2D_{[a}A_{b]\mu\nu}+[A_a,A_b]_{\mu\nu}.
\end{array}
\label{R1}
\end{eqnarray}
Using (\ref{decA}), the latter
%(\ref{R1})
may be
reduced to
\begin{eqnarray}
\begin{array}{lll}
R_{ab\mu\nu}
&=&2\nu_{[b}\{\hat{D}_{a]}\Phi_{\mu\nu}+\dot{A}_{a]\mu\nu}\}\\
&&+2D_{[a}\bar{A}_{b]\mu\nu}+[\bar{A}_a,\bar{A}_b]_{\mu\nu}\\
&&+4\hat{D}_{[a}K_{b][\mu}\delta_{\nu]}^n
  +2K_{a[\mu}K_{\nu]b},
\end{array}
\label{R1B}
\end{eqnarray}
where $\hat{D}_aK_{\mu b}=D_aK_{\mu b}+A_{a\mu}^{\hskip.10in\gamma}
K_{\gamma b}$.
Notice that the intrinsic, three-curvature
$^{(3)}R_{abcd}$ of $\Sigma$
associated with
$h_{ab}$ satisfies
\begin{eqnarray}
\mbox{}^{(3)}R_{ab\mu\nu}=
\bar{D}_a\bar{A}_{b\mu\nu}
-\bar{D}_b\bar{A}_{a\mu\nu}+[\bar{A}_a,\bar{A}_b]_{\mu\nu}.
\end{eqnarray}
It follows that
\begin{eqnarray}
\begin{array}{rl}
h^c_ah^d_bR_{cd\mu\nu}=&
2\bar{D}_{[a}\bar{A}_{b]\mu\nu}+[\bar{A}_a,\bar{A}_b]_{\mu\nu}\\
 &+2K_{a[\mu}K_{\nu]b}+4\hat{\bar{D}}_{[a}K_{b][\mu}\delta_{\nu]}^n\\
=& \mbox{}^{(3)}R_{ab\mu\nu}
+4\hat{\bar{D}}_{[a}K_{b][\mu}\delta_{\nu]}^n\\
 &+2K_{a[\mu}K_{\nu]b}.
\end{array}
\label{R2}
\end{eqnarray}
This last equation generalizes a similar expression obtained in
\cite{sy:a},
where $\mu,\nu\ne n$ is considered.

\subsection{Energy-momentum relations}

%The two reduction formulae (\ref{R1}) and (\ref{R2}) of the Riemann
Constraints related to the initial distribution of energy-momentum
are obtained by consideration of the Ricci tensor.
To this end, consider the two expressions (\ref{R1}) and
(\ref{R2}) for $R_{ab\mu\nu}$
of the previous section in
\begin{eqnarray}
\begin{array}{ll}
h^c_a\nu^bR_{cb\mu\nu}
&=-\hat{D}_a\Phi_{\mu\nu}
-A^\prime_{a\mu\nu},\\
h^c_ah^d_bR_{cd\mu n}&=2\hat{\bar{D}}_{[a}K_{b]\mu}.
\end{array}
\end{eqnarray}
Consequently, the elements of the Ricci tensor become
\begin{eqnarray}
\begin{array}{rl}
\nu^bR_{b\nu}=\nu^bh^{c\mu}R_{cb\mu\nu}\\
=&-\hat{D}^\mu\Phi_{\mu\nu}-A^{\prime\mu}_{\hskip.04in\mu\nu},\\
h^d_bR_{d n}=&2h^d_bh^{a\mu}\hat{\bar{D}}_{[a}K_{d]\mu}\\
=&\bar{D}^aK_{ab}-\bar{D}_bK.
\end{array}
\end{eqnarray}
The second expression (\ref{R2})
for the Riemann tensor also
also gives
\begin{eqnarray}
\begin{array}{rl}
h^\alpha_\mu h^\beta_\nu h^c_a h^d_b R_{ab\alpha\beta}
=&\mbox{}^{(3)}R_{abcd}+K_{a\mu}K_{\nu b}\\&-K_{a\nu}K_{\mu b},
\end{array}
\end{eqnarray}
resulting in the familiar identity
\begin{eqnarray}
\begin{array}{rl}
16\pi T_{nn}=&2(R_{ab}-\frac{1}{2}g_{ab}R)\nu^a\nu^b\\
=&h^{ac}h^{bd}R_{abcd}\\
=&h^{a\mu}h^{b\beta}R_{ab\alpha\beta}\\
=&\mbox{}^{(3)}R+K^2-K_{a\mu}K^{a\mu}\\
=&\mbox{}^{(3)}R+K^2-K_{ab}K^{ab}.
\end{array}
\end{eqnarray}

By Einsteins equations, we have
\begin{eqnarray}
R_{ab}=8\pi(T_{ab}-\frac{1}{2}g_{ab}T)\equiv8\pi\tilde{T}_{ab}.
\end{eqnarray}
Combining the results above, we have the constraints on the
Lie derivative of $A_{a\mu\nu}$ in
\begin{eqnarray}
\hat{D}^\mu\Phi_{\mu\nu}+A^{\prime\mu}_{\cdot\mu\nu}
=-8\pi\tilde{T}_{\nu b}\nu^b,
\label{c_new}
\end{eqnarray}
together with the familiar Gauss-Codacci relations
\begin{eqnarray}
\begin{array}{rl}
\mbox{}^{(3)}R+K^2-K_{ab}K^{ab}&=16\pi T_{nn},\\
\bar{D}^aK_{ab}-\bar{D}_bK&=8\pi h^a_b\tilde{T}_{an}.
\end{array}
\label{c_old}
\end{eqnarray}
We shall refer to (\ref{c_new})-(\ref{c_old})
as the energy-momentum constraints.

In going from a first
order Hamiltonian description to a second order description,
it is not surprising to
encounter also constraints on the Lie derivative,
reflecting the conditions that the Gauss-Codacci
relations also must be satisfied on a future hypersurface
in ${\cal N}(\Sigma)$.

\subsection{Gauss-Riemann relations}

The Gauss-Riemann relations (\ref{GRII}) can similarly be
obtained
in terms of
$\Phi_{\mu\nu}$ and $A_{a\mu\nu}$.
The antisymmetry of the Riemann tensor in its
coordinate indices implies
\begin{eqnarray}
\begin{array}{rl}
\nu^bD^aR_{ab\mu\nu}
=&D^a(\nu^bR_{ab\mu\nu})
-R_{ab\mu\nu}K^{ab}\\
=&D^a(\nu^bR_{ab\mu\nu}).
\end{array}
\end{eqnarray}
By (\ref{R1}), it follows that
\begin{eqnarray}
\begin{array}{rl}
\nu^bR_{ab\mu\nu}
=&-D_a\Phi_{\mu\nu}-\dot{A}_{a\mu\nu}-K_b^{\hskip.04in c}
A_{c\mu\nu}\\
 &-[\omega_a,\Phi]_{\mu\nu}\\
=&-D_a\Phi_{\mu\nu}
-[A_a,\Phi]_{\mu\nu}\\
 &-\dot{A}_{a\mu\nu}-K_b^{\hskip.04in c}A_{c\mu\nu}\\
=&-\hat{D}_a\Phi_{\mu\nu}-A^\prime_{a\mu\nu}.
\end{array}
\end{eqnarray}
Therefore, the
first, inhomogeneous Gauss-Riemann constraint in (\ref{GRII})
reads
\begin{eqnarray}
\begin{array}{rl}
\nu^b\hat{D}^aR_{ab\mu\nu}
=&\nu^bD^aR_{ab\mu\nu}+[A^a,\nu^bR_{ab}]_{\mu\nu}\\
=&D^a(\nu^bR_{ab\mu\nu})
+[A^a,\nu^bR_{ab}]_{\mu\nu}\\
=&\hat{D}^a(\nu^bR_{ab\mu\nu})\\
=&-\hat{D}^a\hat{D}_a\Phi_{\mu\nu}-\hat{D}^aA^\prime_{a\mu\nu}\\
=&\rho_{\mu\nu},
\end{array}
\end{eqnarray}
where the symmetry of the extrinsic curvature
tensor was used in going from the third to the fourth
equation.
These six constraints on
${\cal L}_n A_{a\mu\nu}$ arise in view of the
functional relationship
\begin{eqnarray}
\bar{A}_{a\mu\nu}=\bar{A}_{a\mu\nu}(h_{pq},\partial_rh_{pq}).
\label{frel}
\end{eqnarray}
Because $\frac{1}{2}{\cal L}_nh_{ab}=K_{ab}$ and
$K_{a\mu}=A_{a\mu n}$, six degrees of freedom in
\begin{eqnarray}
{\cal L}_nA_{a\mu\nu}
={\cal L}_n\bar{A}_{a\mu\nu}+2A_{an[\nu}\delta_{\mu]}^n
\end{eqnarray}
are constraint by (\ref{frel}).

To summarize, the constraints on the
initial data in terms of
potentials are\\
\mbox{}\\
{\em{Gauss-Codacci:}}
\begin{eqnarray}
\begin{array}{rl}
\mbox{}^{(3)}R+K^2-K_{ab}K^{ab}=&16\pi T_{nn}\\
     \bar{D}^aK_{ab}-\bar{D}_bK=&8\pi h^a_b\tilde{T}_{an}
\end{array}
\end{eqnarray}
{\em{Lie-constraints:}}
\begin{eqnarray}
\begin{array}{rl}
D^\mu\Phi_{\mu\nu}+A^{\prime\mu}_{\cdot\mu\nu}
=&-8\pi\tilde{T}_{\nu b}\nu^b\\
\hat{D}^a\hat{D}_a\Phi_{\mu\nu}+\hat{D}^aA^\prime_{a\mu\nu}
=&-\rho_{\mu\nu}
\end{array}
\end{eqnarray}
{\em{Lorentz gauge:}}
\begin{eqnarray}
\Phi^\prime_{\mu\nu}+\hat{D}^aA_{a\mu\nu}=0.
\end{eqnarray}

\subsection{Restricted Lorentz gauge}

Restricted gauge transformations
for the connection in analogy to the
same problem in electromagnetics, can be
considered through
the Lorentz transformation $\Lambda_\mu^{\hskip.04in\nu}$
\begin{eqnarray}
\Lambda_\mu^{\cdot\nu}=\delta_\mu^{\cdot\nu}
+\tau\lambda_\mu^{\cdot\nu}
+\frac{1}{2}\tau^2\sigma_\mu^{\cdot\nu}
\mbox{  in   }{\cal N}(\Sigma),
\end{eqnarray}
where $\tau$ is a Gaussian normal coordinate of
$\Sigma$ such that $\nu^c\nabla_c\tau=1$, and
$\partial_\tau\lambda_\mu^{\hskip.04in\nu}=0$.
By (\ref{rule}),
we have
\begin{eqnarray}
\left.
\begin{array}{l}
\Phi^{(r)}_{\mu\nu}=\Phi_{\mu\nu}-\lambda_{\mu\nu}\\
\dot{\Phi}^{(r)}_{\mu\nu}=\dot{\Phi}_{\mu\nu}-\sigma_{\mu\nu}\\
A_{a\mu\nu}^{(r)}=A_{a\mu\nu}
\end{array}
\right\}\mbox{  on  }\Sigma.
\end{eqnarray}
The first (energy-momentum) Lie-constraint
is gauge invariant (invariant under transformations
(\ref{rule})), while the second is not.
Therefore, we can choose
$\lambda_\mu^{\hskip.04in\nu}$ and
$\sigma_\mu^{\hskip.04in\nu}$ so that
\begin{eqnarray}
\left.
\begin{array}{rl}
\Phi^{(r)}_{\mu\nu}&=\Phi_{\mu\nu}-\lambda_{\mu\nu}=0\\
(\Phi^{(r)})^\prime_{\mu\nu}&\equiv
\dot{\Phi}^{(r)}_{\mu\nu}+K\Phi^{(r)}_{\mu\nu}\\
&=\dot{\Phi}_{\mu\nu}-\sigma_{\mu\nu}=\Phi_{\mu\nu}^\prime
\end{array}
\right\}\mbox{  on  }\Sigma.
\end{eqnarray}
The condition on $\sigma_\mu^{\cdot\nu}$
ensures that the second Lie-constraint
remains satisfied.
This shows that by choosing
\begin{eqnarray}
\left.
\begin{array}{l}
\lambda_{\mu\nu}=\Phi_{\mu\nu}\\
\sigma_{\mu\nu}=-K\Phi_{\mu\nu}
\end{array}
\right\}\mbox{  on  }\Sigma
\end{eqnarray}
we obtain
\begin{eqnarray}
\Phi^{(r)}_{\mu\nu}=0\mbox{ on }\Sigma.
\end{eqnarray}
With this restricted Lorentz gauge, the
Lie-constraints
reduce to
\begin{eqnarray}
\left.
\begin{array}{rl}
\hat{D}^aA^\prime_{a\mu\nu}
&=-\rho_{\mu\nu},\\
\dot{\Phi}_{\mu\nu}+\hat{D}_cA^c_{\cdot\mu\nu}&=0.
\end{array}
\right\}\mbox{  on  }\Sigma.
\label{sysYMINR}
\end{eqnarray}

We remark it does not appear to be feasible to ensure
$\Phi=0$ throughout $D^+(\Sigma)$ by suitable
choice of restricted Lorentz gauge on the initial data.

{\bf Acknowledgment.}
The first author greatfully acknowledges the warm hospitality at ITP,
and very stimulating
discussions with S. Chaudhuri, C. Cutler, D. Chernoff,
G. Cook, S.L. Shapiro, S.A. Teukolsky.
The Cornell Theory Center is supported by NSF, NY State, ARPA,
NIH, IBM and others.

\end{document}